\documentclass{article}                  %% LaTeX 2e
\usepackage{opex3}

\usepackage{graphicx}

\newcommand\pictc[5]{\begin{figure}
                          \centering
                       \includegraphics*[width=#1\columnwidth,]{#3}
                   \protect\caption{\protect\label{fig:#4} #5}
                    \end{figure}            }
\newcommand\pict[4][0.7]{\pictc{#1}{!tb}{#2}{#3}{#4}}
\newcommand\rpict[1]{\ref{fig:#1}}

\newcommand\leqt[1]{\protect\label{eq:#1}}
\newcommand\reqtn[1]{\ref{eq:#1}}
\newcommand\reqt[1]{(\reqtn{#1})}

\newcounter{Fig}

\begin{document}
\begin{sloppy}

\title{Incoherent multi-gap optical solitons in nonlinear photonic lattices}

\author{Kristian Motzek$^{1,2}$, Andrey A. Sukhorukov$^1$, Friedemann Kaiser$^2$, and Yuri S. Kivshar$^1$}

\address{$^1$Nonlinear Physics Centre and Centre for Ultra-high
bandwidth Devices for Optical Systems (CUDOS), Research School of
Physical Sciences and Engineering, Australia National University,
Canberra ACT 0200, Australia\\
$^2$Institute of Applied Physics, Darmstadt University of Technology, D-64289 Darmstadt, Germany}

\homepage{http://www.rsphysse.anu.edu.au/nonlinear}

\begin{abstract}
We demonstrate numerically that partially incoherent light can be
trapped in the spectral band gaps of a photonic lattice, creating
partially incoherent multi-component spatial optical solitons in a
self-defocusing nonlinear periodic medium. We find numerically
such incoherent multi-gap optical solitons and discuss how to
generate them in experiment by interfering incoherent light beams
at the input of a nonlinear periodic medium.
\end{abstract}


\section{Introduction}

Gap solitons are nonlinear localized waves which are associated
with band-gaps in the transmission spectra of nonlinear photonic
structures with a periodically modulated refractive
index~\cite{Voloshchenko:1981-902:ZTF, Chen:1987-160:PRL,
deSterke:1994-203:ProgressOptics}, such as fiber Bragg
gratings~\cite{Eggleton:1996-1627:PRL}, waveguide
arrays~\cite{Mandelik:2003-53902:PRL, Mandelik:2004-93904:PRL},
and optically-induced lattices~\cite{Fleischer:2003-147:NAT,
Fleischer:2003-23902:PRL, Neshev:2004-83905:PRL}. Such gap
solitons are composed of counter-propagating waves which
experience Bragg reflection from the periodic structure and remain
trapped at the nonlinearly-induced lattice defect. It was recently
predicted theoretically~\cite{Cohen:2003-113901:PRL,Sukhorukov:2003-113902:PRL,
Pelinovsky:2004-36618:PRE} and demonstrated experimentally~\cite{Hanna:2004:ProcNLGW} that gap solitons can be
composed of many modes which are localized in multiple band-gaps
of a single periodic lattice. When several modes are excited
simultaneously with a coherent light source, they can trap
together in the form of a periodic or quasi-periodic multi-band
breather~\cite{Mandelik:2003-253902:PRL}. On the other hand,
stationary multi-gap solitons with fixed profile along the
propagation direction may form if the modes are made mutually
incoherent, and non-instantaneous nonlinear response is defined by
the time-averaged light intensity. It was suggested
theoretically~\cite{Buljan:2004-223901:PRL} that multi-gap
solitons can be generated by a partially coherent light source
which simultaneously excites multiple modes in different gaps of
the lattice spectrum.

A multi-gap soliton can be generated in {\em a self-focusing
nonlinear periodic medium} by a single partially coherent beam
incident at normal angle on a periodic photonic structure, as was
recently demonstrated in experiment for optically-induced photonic
lattices~\cite{Cohen:2005-500:NAT}. The self-trapping mechanism of such
a multi-gap optical soliton is mainly associated with the index
guiding due to the total internal reflection experienced by the
strongest modes, which support much weaker guided modes in the
Bragg-reflection gaps of the spectrum. However, from the general
concept of the multi-gap
solitons~\cite{Cohen:2003-113901:PRL, Sukhorukov:2003-113902:PRL, 
Pelinovsky:2004-36618:PRE} it follows that such
multi-component solitons can also exist in {\em self-defocusing
nonlinear periodic media}, but this is only possible when all the
modes are localized in the Bragg-reflection gaps of the lattice
spectrum.

In this paper, we demonstrate numerically that partially
incoherent multi-gap solitons can indeed exist in self-defocusing
nonlinear media, and we suggest a simple approach for creating an
input light field with special coherence properties, which allow
for excitation of the modes in particular gaps, and highly
efficient generation of incoherent multi-gap solitons.

\section{Model and multi-gap optical solitons}

Propagation of partially coherent light beams through a nonlinear
medium with a slow response can be described by a nonlinear
parabolic equation for the envelope $A$ of the electric field,
\begin{equation} \leqt{parabolic}
  i \frac{\partial A}{\partial z} + D \frac{\partial^2 A}{\partial x^2} +
  k_0^2[\nu^2(x)+\delta n^2(I)] A = 0 \, ,
\end{equation}
where $z$ and $x$ are the propagation and transverse coordinates,
respectively, $D = 1/(2 k_0 n_e)$ is the diffraction coefficient,
$k_0$ is the vacuum wavenumber, $n_e$ is the average refractive
index of the unperturbed crystal, $\nu(x)$ is the periodic
modulation of the refractive index, and $\delta n(I)$ is the
nonlinearity-induced change of the refractive index that depend on
the time-averaged light intensity $I$. We consider the case of
slow {\em self-defocusing nonlinearity}, $\delta^2 n(I) \simeq -
\gamma I$, where $\gamma > 0$ is the nonlinear coefficient, and we
neglect the effect of the nonlinearity saturation.

\pict[0.65]{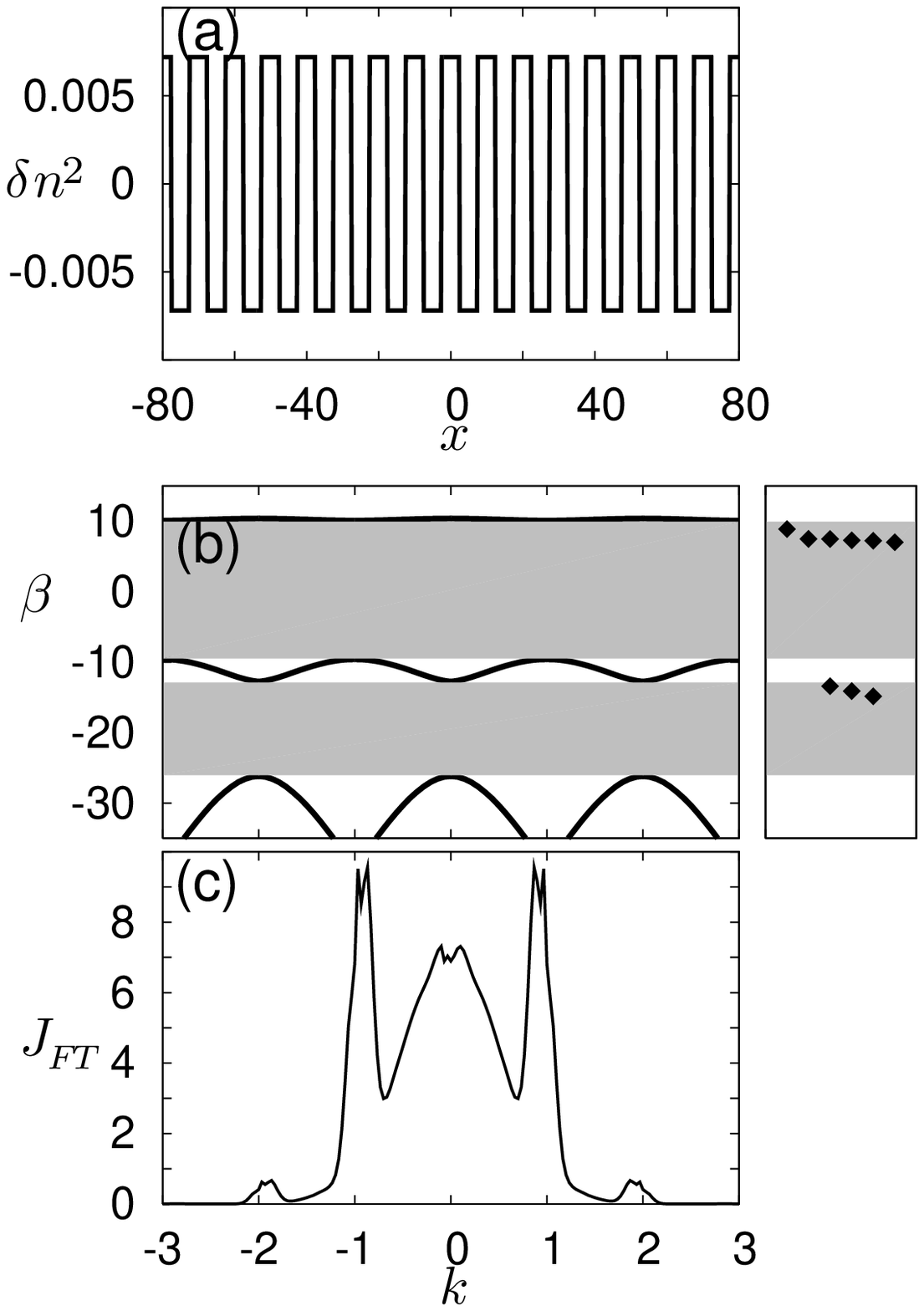}{bands}{
(a) Refractive index and (b) structure of the bandgap
spectrum of the photonic lattice. The first and second gap are
shaded. Right panel shows the propagation constants of the
different components of the multi-gap soliton shown in
Fig.~\rpict{soliton}. Six of them lie in the first gap, and three
in the second band-gap. The power spectrum of the soliton, found as a sum of Fourier power spectra of individual components, is shown in (c). }

For a partially coherent light source, the phase of the light
field envelope $A(x,t)$ changes very rapidly (i.e. on a time scale
much faster than the response time of the crystal). There exist
several equivalent approaches to analyze such an incoherent light
theoretically. The coherent density approach
\cite{Christodoulides:1997-646:PRL} and the self-consistent
multimode theory~\cite{Mitchell:1997-4990:PRL} are based on the
fact that incoherent light can be decomposed into the components,
$A(x,t)=\sum_m \phi_m(x)\exp[i\gamma_m(t)]$, which are mutually
incoherent due to the random phase factors $\gamma_i(t)$.

In this paper, we follow Ref.~\cite{Buljan:2004-223901:PRL} and
employ the self-consistent multimode theory to find
multi-component solutions for incoherent gap solitons. The
refractive index distribution associated with such a soliton is
created by several localized modes. Because the soliton profile
does not change under propagation, all the functions $\phi_m(x)$
should correspond to the modes of the soliton-induced waveguide.
These may include both fundamental and higher-order modes localized in multiple band-gaps. %---endnew
To find incoherent solitons numerically, we use the following
procedure. First, we start with a refractive index distribution and
calculate, by means of linear algebra, all localized and radiation modes of
this refractive index distribution, using Eq.~\reqt{parabolic}. 
Second, we construct the solutions based on a set number of lowest order localized modes in the first one or two Bragg-reflection band-gaps. We scale the mode amplitudes to match the chosen intensity and power levels, and calculate the time-average refractive index distribution for the multi-mode light field. Then, we repeat the calculations again until a self-consistent multi-component solution is reached with a certain accuracy.

\pict[0.7]{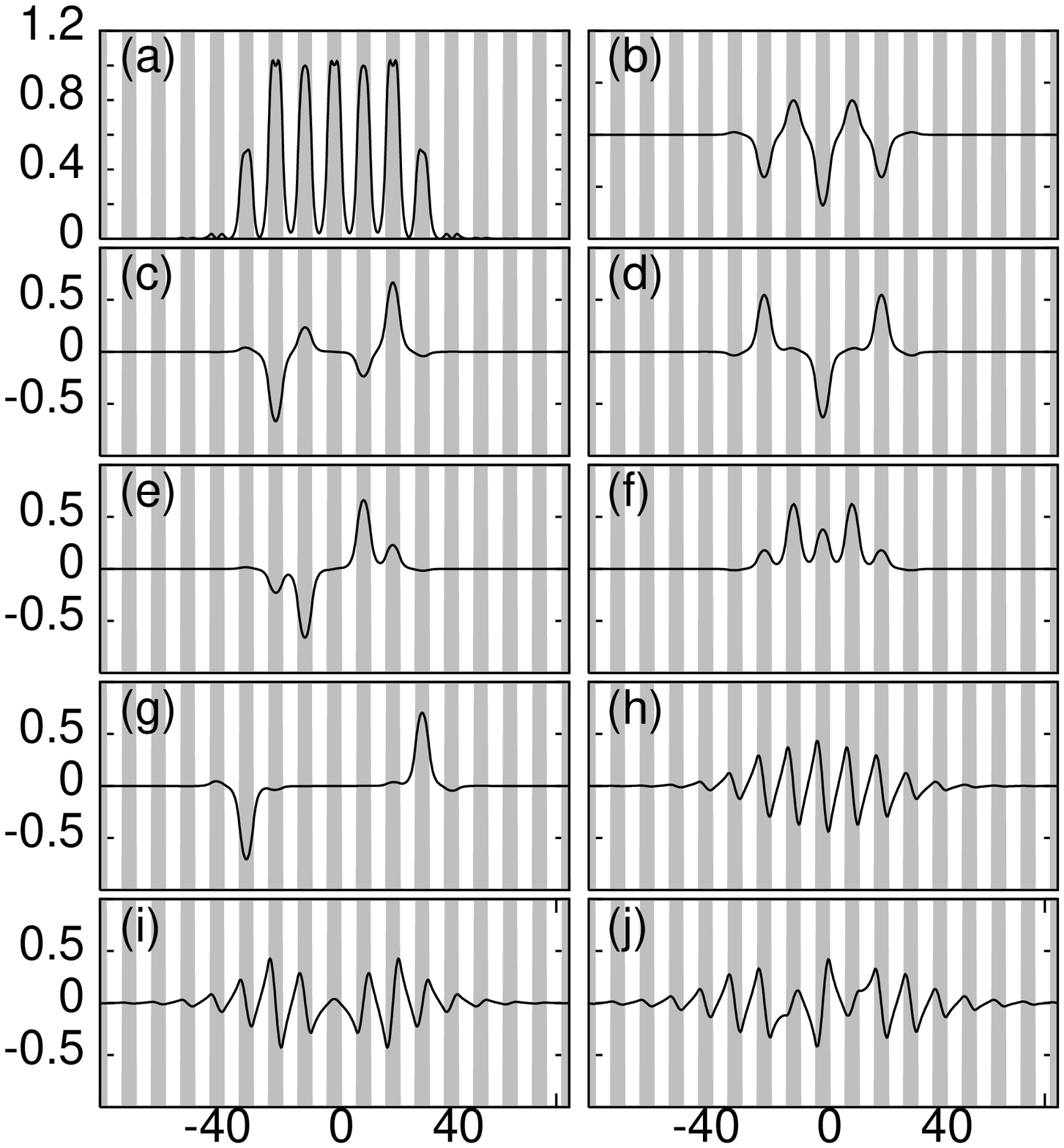}{soliton}{
The numerically obtained incoherent gap soliton. Shown
are (a)~the total soliton intensity, and its separate components
in (b-g) the first gap and (h-j)~second spectral gap. }

For our study, we choose a step-like modulation of the refractive
index with a period of $10 \mu$m and varying in the value between
$\nu_+=7.20\times 10^{-3}$ and $\nu_-=-\nu_+$. As for other
parameters, we assume light of a vacuum wavelength of $532$nm and
an unperturbed refractive index $n_e=2.3$. The bandgap spectrum of
such a periodic structure is shown in Fig.~\rpict{bands}. Note
that we choose a rather strong modulation of the refractive index
(\lq a deep lattice\rq). This results in a considerable width of
the first two band gaps.
The third band gap, however, is still rather narrow and can only support weakly localized modes. In the following, we consider solitons based on strongly localized modes in the first and second gaps. 

\pict[0.75]{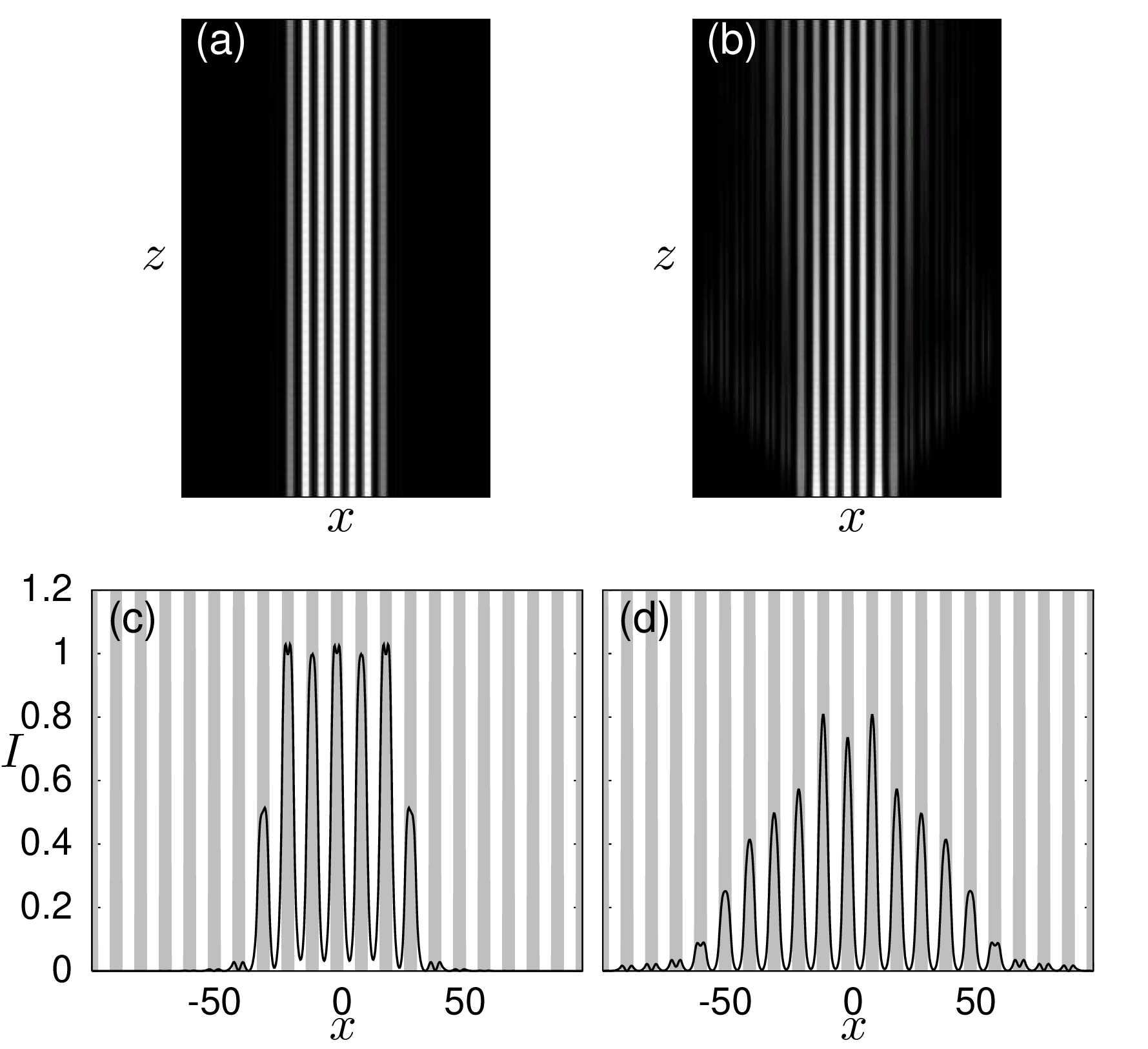}{propag}{
(a,b) Propagation of an incoherent gap soliton for 20mm
in the nonlinear and linear regimes, respectively. The
corresponding output beam profiles are shown in (c,d).}

Using the self-consistent multimode theory, we find the incoherent
gap solitons numerically. We were able to find solitary solutions
containing only localized modes from the first gap, but also
solutions where a considerable amount of the light intensity is
trapped in the second gap. In Fig.~\rpict{bands}, we show an
example of that second type presented by the propagation constants
of the localized modes that form the incoherent gap soliton. The
soliton consists of six modes localized in the first gap and three
modes localized in the second gap.

Figure~\rpict{soliton} shows the profile of the incoherent gap
soliton and the corresponding set of single modes that create the
soliton.
The components from the first gap [Figs.~\rpict{soliton}(b)-(g)] were
chosen to have equal power, the components from the second spectral gap
[Figs.~\rpict{soliton}(h)-(j)] were chosen to have 75\% of that power. The
power was chosen such that we obtained a soliton with a peak intensity of around 1. The resulting propagation constants are shown in Fig.~\rpict{bands}(b). For components from the first gap [Figs.~\rpict{soliton}(b)-(g)], the corresponding propagation constants are 8.89284, 7.48452, 7.47812, 7.29089, 7.22568, 7.02095, and in the second gap [Figs.~\rpict{soliton}(h)-(j)] the values are -13.4242, -14.1175, -14.8541.

We notice two remarkable differences comparing to the
case of {\em coherent gap solitons} studied earlier. The first
feature of the incoherence is that the gap soliton is very broad
with a rather flat top. The soliton has to be broad, otherwise it
could not support so many localized modes, especially in the
second bandgap where localized modes tend to have a large spatial
extent. The second feature introduced by the incoherence is that
there is some light intensity even in the regions of lower
refractive index.
Near the center of the soliton the lowest intensity in the lower index 
region is still about 3-4\% of the maximum intensity of the soliton. Although
this might be a small effect, it is still a qualitative difference to the
coherent case, where the intensity inevitably drops to zero somewhere in the
lower index region. We note that coherent gap solitons can form stable bound states for a particular phase difference~\cite{Kartashov:2004-2831:OE}, however the interacting properties of multi-mode gap solitons are expected to be more complex due to nontrivial effect of coherence~\cite{Ku:2005-63904:PRL}.

In order to verify that the incoherent gap soliton is stable and
that the nonlinearity is essential to keep it localized, we have
numerically propagated the soliton, once with the nonlinearity
switched off and once with the nonlinearity switched on, but in
the presence of an initial perturbation. The results are
summarized in Figs.~\rpict{propag}(a-d). We observe that the
initial perturbation does lead to internal oscillations of the
soliton as it propagates, and the soliton retains its structure,
demonstrating robustness under the action of perturbations.
However, when the nonlinearity is switched off, we observe that
diffraction sets in. Due to the deep lattice using for this
example, the light only diffracts slowly. However, the length
scales are still small enough to make possible experiments to
check these numerical results seem feasible.

\section{Generation of incoherent gap solitons}

\pict[0.5]{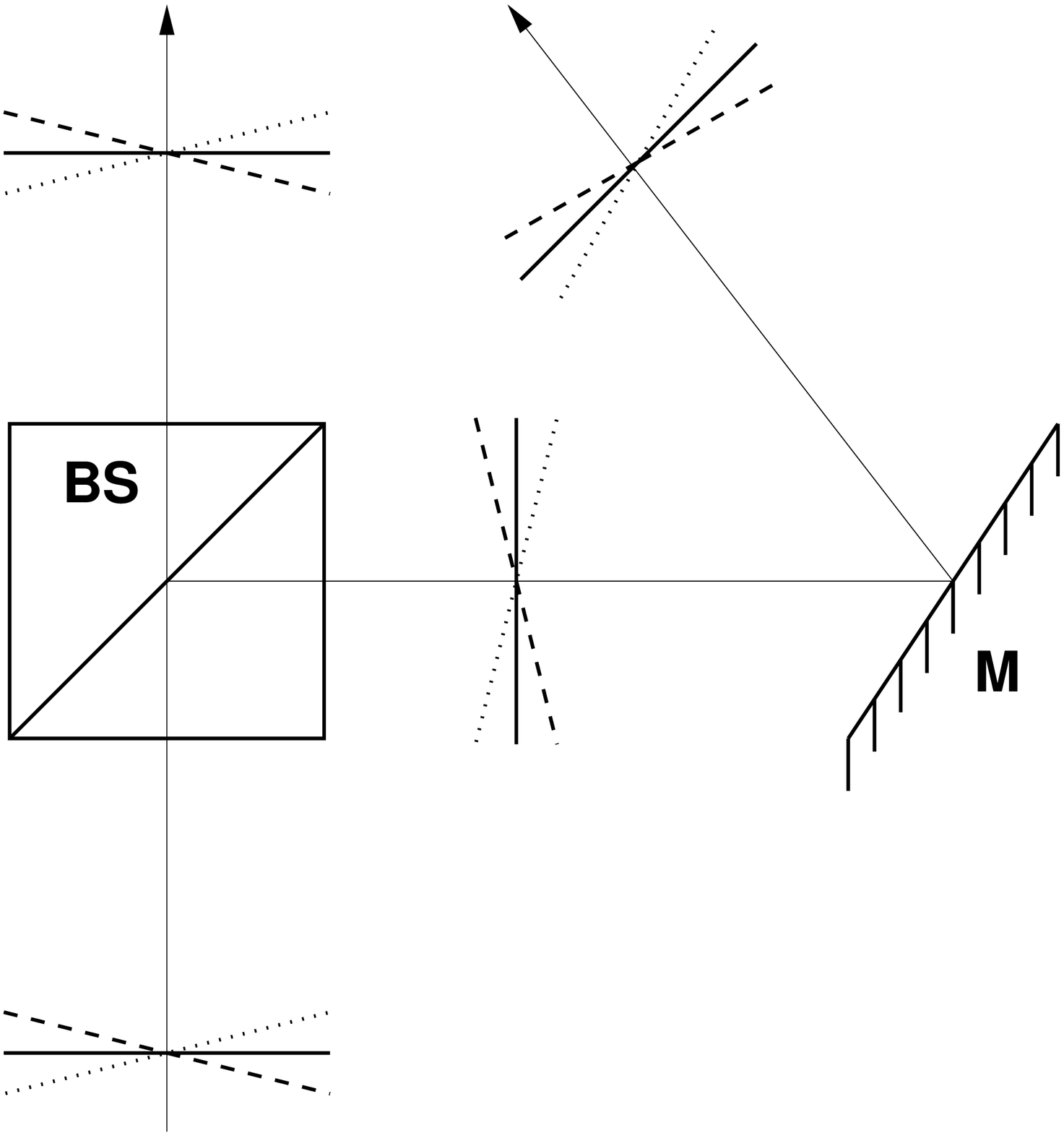}{scheme}{
Schematic for the generation of multi-gap solitons.
First, an input partially incoherent beam splits into two beams by
a beam splitter and then two beams are recombined at the input of
the nonlinear periodic medium. Solid, dashed, and dotted lines
stand for the mutually incoherent plane waves that partially
incoherent light can be decomposed into. The plane waves interfere
at the input face of the nonlinear medium.}

We now discuss the approaches which can be used to generate
partially coherent gap solitons in self-defocusing nonlinear
media. According to the coherent density method, a uniform
incoherent light beam can be regarded as a superposition of many
mutually incoherent plane waves that are all tilted at small
angles with respect to each other. In order to realize the
efficient generation of stationary multi-gap solitons, one has to
selectively excite only the Floquet-Bloch modes corresponding to
the upper gap edges, which experience {\em anomalous diffraction}
in the periodic medium. Since the Bloch waves are composed of
counter-propagating waves with a particular phase difference,
their controlled excitation can be performed by a partially
coherent input, where all pairs of waves propagating at the angles
$\alpha+\alpha_B$ and $\alpha-\alpha_B$ are fully correlated, and form
interference patterns matching the Bloch-wave profiles (here $\alpha_B$ is the
Bragg angle).
The angle $\alpha$ is introduced, because the interference patterns do not
necessarily have to propagate parallel to the $z$-direction. Hence, $\alpha$
is the angle between the direction of propagation of the interference pattern
and the $z$-axis. Such angles need to be considered, since incoherent light
can be most conveniently described mathematically as an incoherent
superposition of fully coherent components, all propagating in slightly
different directions.
These requirements can be satisfied by using a simple set-up illustrated in
Fig.~\rpict{scheme}. First, a light beam is made partially
incoherent, e.g. by passing a laser light through a rotating
diffuser~\cite{Segev:2001-87:SpatialOptical}. Then, the input beam
is split into two beams, which are now fully correlated with
respect to each other. If optical paths of these beams are made
equal until they are incident on the crystal under the angles
$\theta$ and $-\theta$, and additionally the beams exhibit
different number of reflections, then the components with the
opposite propagation angles will be fully correlated, so that the
light amplitude at the input face can be presented in the form
\begin{equation}
\leqt{interf}
  A(x,t)=\sum_j G(\alpha_j) e^{ik_0(z+\alpha_jx)}
  \cos[k_0\theta (x-x_i)]e^{i\gamma_j(t)}.
\end{equation}
We thus have a superposition of interference patterns with the
same lattice constant $d_i = (k_0 \theta)^{-1}$ but propagating at
slightly tilted angles $\alpha_j$. These patterns will excite
Bloch waves at a particular band edge which symmetry is selected
by a choice of the parameters $d_i$ and $x_i$. The suggested
generation scheme provides a generalization of the approach which
was recently used for generating coherent spatial gap
solitons~\cite{Mandelik:2004-93904:PRL, Neshev:2004-83905:PRL}.
It is also reminiscent of experimental setup which was recently used to investigate the interaction between partially incoherent solitons~\cite{Ku:2005-63904:PRL}. Note, however, that in contrast to those experiments, the beams have to fully overlap on the input face of the medium in our case and that they have to do so under a certain angle.

\pict[0.75]{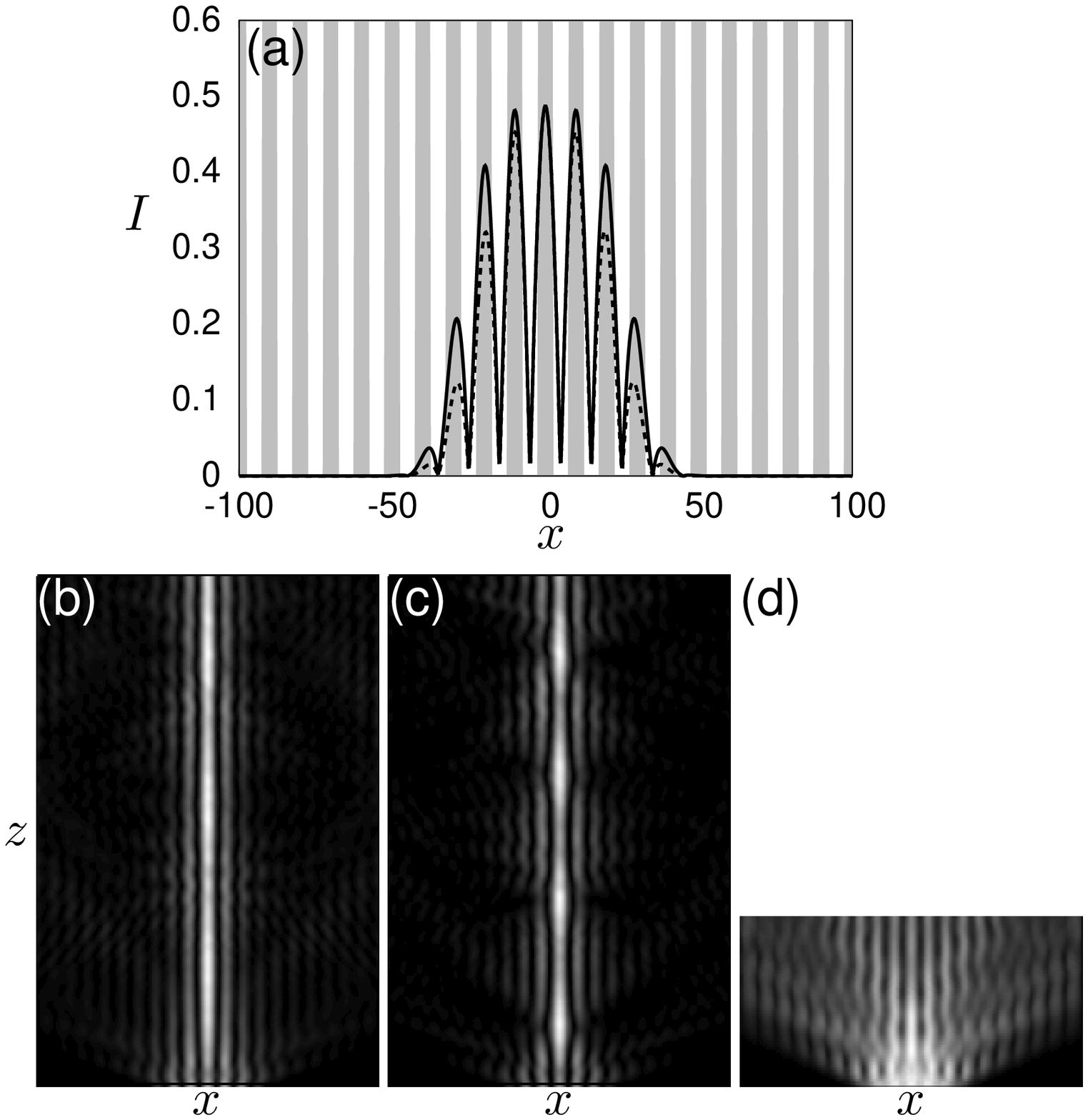}{generation}{
Example of the generation of an incoherent multi-gap
soliton. (a)~Output (dashed) and input (solid) intensity profiles
corresponding to case (b); (b-d)~intensity patterns along the
propagation direction for input excitation by inclined overlapping beams with the same profiles but different coherence properties: (b)~partially incoherent,
correlated counter-propagating Fourier components; (c)~fully
coherent; (d)~partially incoherent, uncorrelated Fourier
components. }

To illustrate the generation of the gap solitons with multiple
modes trapped in the first gap, we perform numerical simulations
and select the following parameters: $d_i = d/2$ and $x_i = 0$.
The soliton dynamics is summarized in Fig.~\rpict{generation},
where we use a shallow lattice ($\nu_+=-\nu_-=1.2 \times 10^{-3}$) to make all effects more visible. The input and output (after $16$mm) intensity profiles are
presented in in Fig.~\rpict{generation}(a). We observe highly
efficient generation of immobile multi-gap soliton with minimal
amount of radiation, in a sharp contrast to the case of a
homogeneous self-defocusing medium.

What happens if only the power spectrum of the soliton is matched
at the input, for example, by performing Fourier filtering of a
single partially-coherent beam, whereas the phases of waves with
opposite propagation angles remain uncorrelated? In this case, the
Bloch waves corresponding to the top and bottom gap edges would be
excited simultaneously, however they experience self-trapping or
enhanced broadening in self-defocusing media, respectively.

We perform numerical simulations using the same overlapping beams at the input as in
Fig.~\rpict{generation}(b), but making counter-propagating waves
{\em uncorrelated}, see Fig.~\rpict{generation}(d). Although the
power spectrum at the input remains almost the same, the interference
fringes disappear, and the whole beam diffracts strongly due to
self-defocusing of the modes at the bottom gap edge which carry
about 50\% of the beam power. We note that soliton generation may be observed under similar conditions in a very deep lattice, where the band-1 modes diffract much slower than the higher order bands, however substantial radiation losses may still occur.

On the other hand, if the input is made fully coherent, strong focusing is observed leading to the formation of a breathing state, as shown in
Fig.~\rpict{generation}(c). These results illustrate the
importance to engineer the coherence properties of the input light
for the efficient generation of gap solitons especially in shallow lattices, in contrast to the random-phase lattice solitons~\cite{Cohen:2005-500:NAT} where such engineering is not required.

\section{Conclusions}

We have demonstrated that, in self-defocusing media with a
periodically modulated refractive index, partially incoherent
light can be trapped in the gaps of the structure bandgap
spectrum, and it can even propagate in the form of bright
incoherent gap solitons. The interplay between the nonlinearity
and the lattice discreteness is essential for this to be possible.
Furthermore, we have shown that it should be possible to generate
such incoherent gap solitons in experiments by splitting an
incoherent plane wave into two beams and then recombining these
beams under an appropriate angle at the input of a nonlinear
periodic medium.

\end{sloppy}
\end{document}